\documentclass[amsmath,amssymb,twocolumn]{revtex4}
\usepackage[T1]{fontenc}
\usepackage{graphicx}
\usepackage{subfig}
\usepackage{epsfig}
\usepackage{dcolumn}
\usepackage{multirow}
\usepackage{rotating}
\usepackage{verbatim}
\usepackage{bm}
\usepackage{color}
\usepackage{soul}
\usepackage{float}
\usepackage{hyperref}
\usepackage[utf8]{inputenc}

\def\lsim{\raise0.3ex\hbox{$<$\kern-0.75em\raise-1.1ex\hbox{$\sim$}}}
\def\gsim{\raise0.3ex\hbox{$>$\kern-0.75em\raise-1.1ex\hbox{$\sim$}}}

\def\beqa{\begin{eqnarray}}
\def\eeqa{\end{eqnarray}}

\begin{document}

\title{Investigating the inclusive transverse spectra in high-energy $pp$ collisions in the context of geometric scaling framework}
\author{L. S. Moriggi}
\email{lucas.moriggi@ufrgs.br}
\author{G.M. Peccini} 
\email{guilherme.peccini@ufrgs.br}
\author{ M.V.T. Machado}
\email{magnus@if.ufrgs.br}
\affiliation{High Energy Physics Phenomenology Group, GFPAE. Institute of Physics, Federal University of Rio Grande do Sul (UFRGS)\\
Caixa Postal 15051, CEP 91501-970, Porto Alegre, RS, Brazil}

\begin{abstract}

The presence of geometric scaling within the $p_T$ spectra of produced hadrons at high energy $pp$ collisions using small-$x$ $k_T$-factorization is investigated. It is proposed a phenomenological parametrization for the unintegrated gluon distribution in the scaling range that reproduces the features of the differential cross section both in the saturated and dilute perturbative QCD regimes. As the saturation scale acts as an effective regulator of the infrared region, the extension of the model to quantities usually associated to soft physics is studied. The approach is applied to compute the average $p_T$ and the rapidity distribution of produced gluons at high energies.
\end{abstract}

\maketitle

\section{Introduction}
The transverse momentum spectra of produced hadrons in  $pp(\bar p)$ collisions is an observable that has been analyzed in different experiments, from fixed target ones at low energies up to LHC energies. It is well known that the features of this spectrum can be reproduced by a function that has a powerlike falloff with a power index, $n$,  at large $ p_T $, while the semihard region of moderate or small $ p_T $ depends on a relative momentum scale, $p_0$. These characteristics can be represented by the function of Hagedorn \cite{Hagedorn}, $ \frac{Ed ^3\sigma} {d^3\vec{p}}=C\left(1+ \frac {p_T} {p_0}\right)^{-n}$, which may also be interpreted as a Tsallis distribution \cite{Tsallis:1987eu}. Thus, the parameters have their  meaning in the context of nonextensive statistical mechanics, with $n$ associated with entropy and $ p_ {0} $ related to the temperature. Phenomenological fits based on Tsallis distribution have shown great precision in describing data from different colliders over a wide range of collision energies, $\sqrt{s}$ \cite{Marques}. The hard-scattering of pointlike particles predicts an index $n=4$, while perturbative corrections generate a rise in this value. It is higher at lower $\sqrt{s}$, i.e. $n\simeq 8$,  and close to $n=6$ for collisions at TeV scale \cite{Brodsky:2005fza,Arleo:2009ch}. The collinear factorization framework in perturbative QCD (pQCD) predicts an effective rise of this index due to the resummation of terms containing powers of $\alpha_s\log{Q^2}$ in the cross section associated with the emission of collinear radiation. Therefore, the value of the parameter $n$ is directly connected to the dynamics of the partonic distributions and the QCD factorization at hard momentum scales. The presence of a typical momentum scale that determines the growth of the cross section at high energies and at small $p_T$ is predicted within the saturation/Color Glass Condensate (CGC) framework, i.e. the saturation scale, $Q_s(x)$. This quantity establishes the region in which the gluon distribution has its maximum value, resulting in a slower growth of the cross section above that limit. This behavior emerges in the data through the geometric scaling on the variable $\tau=Q^2/Q_s^2(x)$, indicating that the cross section does not depend separately on $ Q^2$ and $x$ but rather on the ratio between the momentum and saturation scales. This phenomenon has been reported in different observables \cite{Stasto,Goncalves:2003ke,Armesto:2004ud,Marquet:2006jb,Ben:2017xny}, even in the regime of relative high momentum, $Q^2 \gg Q_s^2(x)$, which would imply that the parameter $n$ related to the $p_T$ spectra should be a function of the scaling variable within this regime. It can be traced back to the geometric scaling behavior of the unintegrated gluon distribution (UGD) in both the target and projectile at sufficiently high energies. This is the main guidance in phenomenological analysis that we will perform in the present work.

While the collinear factorization framework is well established to calculate observables at high $Q^2$, the saturation framework makes use of $k_T$-factorization at small-$x$ regime and semihard momentum scales where the fundamental quantity is the UGD, $\phi(x,k_T^2)$, which is transverse momentum dependent and directly related to the QCD color dipole cross section, $\sigma_{q\bar q}(x,r)$. Distinct approaches have been proposed to model the dipole cross section \cite{GBW1,GBWnovo,IIM,DHJ,BUW,AAMQS,AGL1} and although they give close results in the saturated regime, at the limit of small dipole sizes, $r$, different behaviors for the large $k_T$ tail of the gluon distribution \cite{Machado:2005ez} is predicted. This fact leads to  large differences in the $p_T$ spectra at $p_T > Q_s(x)$. In this work we investigate the presence of geometric scaling in inclusive hadron production using the $k_T$-factorization approach, where a parametrization for an UGD which could be more directly related to the $p_T^{-n} $ behavior at large $p_T$, is proposed. We also discuss the role played by the hadronization process regarding the transition of produced gluons into hadrons and how the scaling would be violated in this situation. Within the saturated regime, $p_T<Q_s(x)$, the saturation framework has the advantage that the saturation scale regulates the typical divergent infrared (IR) behavior of the cross sections, which gives us the possibility to calculate observables usually associated with nonperturbative physics like the total $pp$ cross section \cite{Bartels,guilherme,Carvalho:2007cf}. Furthermore, we analyze the feasibility  of our  parametrization in describing  observables that involve the soft region, such as the rapidity distribution and the average $p_T$ of the produced gluons. This paper will be organized as follows. In Sec. \ref{sec:model} we present the model for the UGD based on general aspects of hadron $p_T$ spectra. Predictions for the invariant cross section and averaged $p_T$ are provided as well as a new geometric scaling parametrization for Deep Inelastic Scattering (DIS) cross section in the small-$x$ region. In Sec. \ref{sec:results}, the results are compared against experimental data of DIS and $p_T$ hadron spectra of neutral and charged particles. Predictions for the rapidity distribution and mean transverse momentum of produced gluons are also shown. Finally, in Sec. \ref{sec:conc} we summarize the main points and results and expose our conclusions.

\section{Theoretical framework and main predictions}
\label{sec:model}
In the color dipole approach applied to DIS, the virtual photon is decomposed by its hadronic Fock states, which in leading order (LO) are a quark-antiquark pair, $q\bar{q}$. The interaction with the target is described in two stages: the fluctuation of the virtual photon into the $q\bar{q}$ pair and its subsequent interaction with the hadronic target. Concerning the first stage, the probability of the virtual photon fluctuating  into the $q\bar{q}$ (with $z$ and ($1-z$) being the longitudinal momentum fraction of the quark and the antiquark, respectively) is given by the photon wave function squared, $ |\Psi(z,r)|^2$, where $r$ stands for the transverse size of separation between the quark and the antiquark. In the second stage, the interaction between the dipole and the target is computed by the dipole cross section $\sigma_{q \bar q}(r)$, \cite{Nikolaev1,Nikolaev2},
\begin{eqnarray}
\label{eq:DISa}
\sigma_ {(L,T)}^{\gamma^* p}(x,Q^2) &= & \int_0^1 dz \int d^2r |\Psi _{(L,T)} (r, z)| ^2 \sigma_{q \bar q}(x,r), \\
|\Psi_L(z,r)|^2    &= & \frac{6\alpha_{em}}{(2\pi)^2}\sum_{n_f}4e_f^2Q^2z^2(1-z)^2K_0^2(\epsilon r), \nonumber \\
|\Psi_T(z,r)|^2   & = &\frac{6\alpha_{em}}{(2\pi)^2}\sum_{n_f}e_f^2\biggl\{  [z^2+(1-z)^2]\epsilon^2 K_1^2(\epsilon r) \nonumber \\
                           &+   &     m_f^2K_0^2(\epsilon r)   \biggr\} ,
\end{eqnarray}
where $\epsilon^2 = z(1-z)Q^2+m_f^2$ and $K_\nu$ are the Modified Bessel Functions of the second kind. The summation over the quark flavors with masses $m_f$ and charges $e_f$ is explicitly shown.  Following the optical theorem, we can determine the dipole cross section considering that the impact parameter dependence is factorized, that is,
\begin{eqnarray}
\label{eq:DISb}
\sigma_{q \bar q}(x,r)=2\int d^2b [1-S(x,r,b)]=\sigma_0[1- S(x,r)].
\end{eqnarray}
In the expression above, $S(x,r)$ is the dipole scattering matrix, and $\sigma_0 =2\pi R_p^2$ is twice the proton transverse area. It was assumed a Heaviside function for the impact parameter dependence, $S(x,r,b)=S(x,r)\Theta (R_p-b)$.  At the limit of large dipoles, $S(x,r) \rightarrow 0$, and the dipole cross section reachs its maximum, $\sigma_0$. The unintegrated gluon distribution function can be obtained from the Fourier transform of the dipole cross section \cite{Barone1,Nikolaev:1994ce},
\begin{eqnarray}
\label{dip-ugd}
\sigma_{q\bar q}(x,r)= \frac{4 \pi}{3} \alpha_s \int \frac{d^2k_T}{k_T^2}(1-\exp(i\vec{k_T}\cdot \vec{r})) \phi(x,k_T^2).
\end{eqnarray}

The cross section for inclusive gluon production with transverse momentum $p_T$ and rapidity $y$ shall be calculated using the $k_T$- factorization approach \cite{Gribov:1983fc},
\begin{eqnarray}
\label{eq:fatkt}
E\frac{d^3\sigma}{dp^3}^{ab \rightarrow g+X}=\frac{2\alpha_s}{C_F}\frac{1}{p_T^2}\int d^2k_T\phi(x_a,k_T^2)\phi(x_b , (p_T-k_T)^2),
\end{eqnarray}
where $x_{a,b}$ are the forward light cone variables of colliding partons (gluons), respectively. That is, 
\begin{eqnarray}
x_a=\frac{p_T}{\sqrt{s}}e^y, \qquad x_b=\frac{p_T}{\sqrt{s}}e^{-y}. 
\end{eqnarray} 

\begin{figure}[t]
\includegraphics[width=\linewidth]{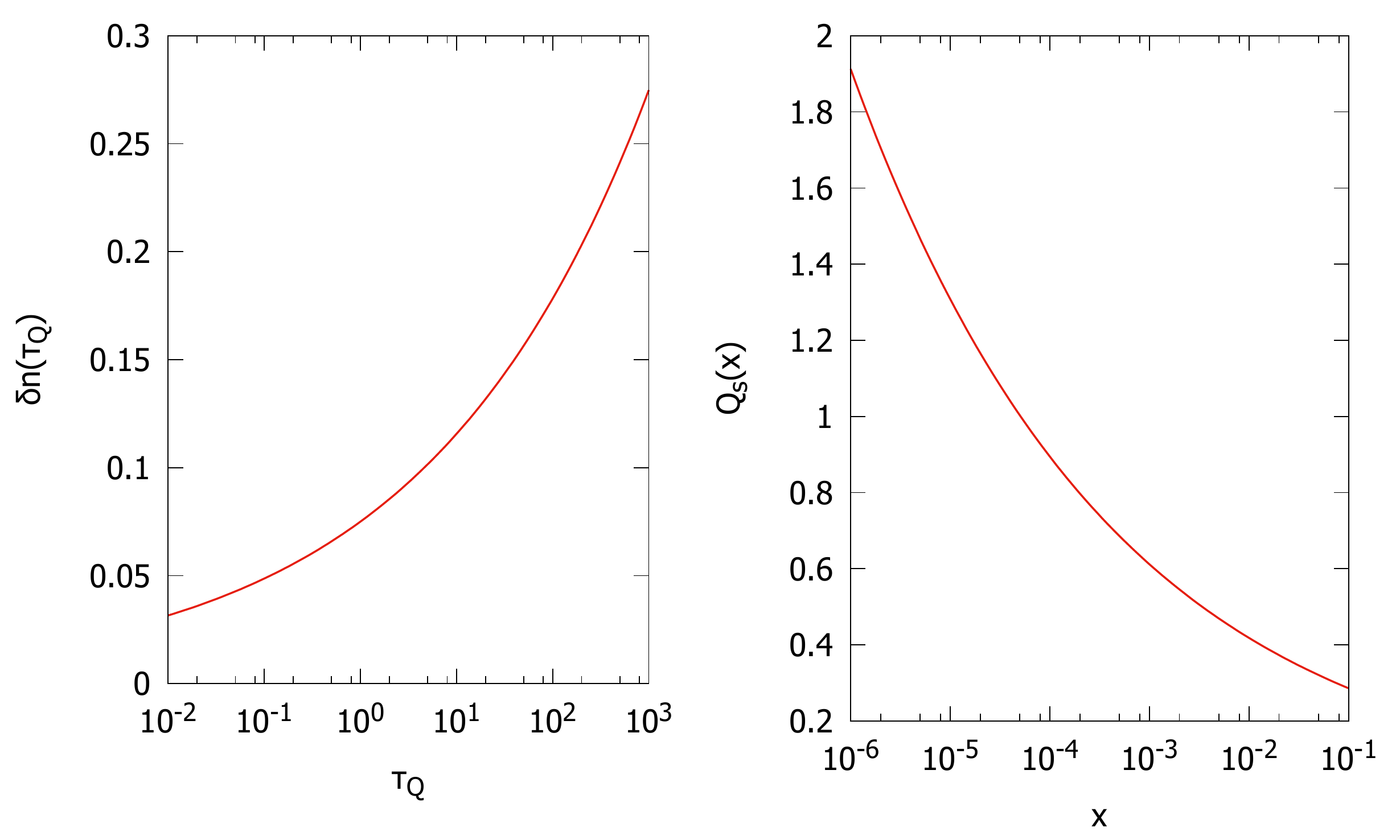}
\caption{$Q_s(x)$ and $\delta n (\tau)$ obtained from parametrization of Eq. (\ref{eq:pars}), with parameters determined from HERA data.}\label{fig:parametros}
\end{figure}

In Eq. (\ref{eq:DISa}), the saturation effects on the gluon distribution leads to the $\gamma^*p$ cross section remaining finite as $Q^2 \rightarrow 0$. The saturation scale works as a regulator for the soft region without the need of  {\it ad hoc} cutoff. On the other hand, the cross sections for jet production show a divergent behavior in the IR region when $p_T$ goes to zero, which is an important feature of perturbative interactions. If we analyze Eq. (\ref{eq:fatkt}), the divergence in $p_T^{-2}$ when $p_T \rightarrow 0$ can be clearly verified. This divergence is smoother than the one from the parton model, namely $\sim p_T^{-4}$ divergence. Nevertheless,  a cutoff is still needed. The authors in Refs. \cite{Levin:Rezaeian,Tribedy,Gay2} raised the possibility of implementing the regularization through the mass of the produced jet, $p_T^2 \rightarrow p_T^2+m^2$. Analogously to the case of minijet models, this type of cutoff requires a mass scale that increases with energy, leading to the presence of two dependent energy scales in the small $p_T$ region. Another possibility has been addressed in \cite{KLN1,KLN2}, in which the authors rewrite Eq. (\ref{eq:fatkt}) separating it into two regions of integration: $k_T\ll p_T$ and $|\vec{p}_T-\vec{k}_T| \ll p_T$. Thereby, as $p_T$ goes to zero the integral also vanishes without any dependence on a new momentum scale. Thus, the expression for the invariant cross section can be approximated as
\begin{eqnarray}
\label{eq:fatkt2}
E\frac{d^3\sigma}{dp^3}^{ab \rightarrow g+X}&= &\frac{\alpha_s}{C_F}\frac{1}{p_T^2}\biggl\{    \phi(x_a,p_T^2)\int^{p_T^2} d^2k_T\phi(x_b , k_T^2)   \nonumber \\
         &  + & \phi(x_b,p_T^2)\int^{p_T^2} d^2k_T\phi(x_a , k_T^2)  \biggr\}. 
\end{eqnarray}

Now, we will introduce the main point in the present work. A simple way to incorporate this behavior within gluon distributions in IR region is to consider an effective regulator  of the gluon propagator compatible with a Yukawa potential, $\phi(k_T^2)\sim \alpha_s  k_T^2/(1+k_T^2/\mu ^2)$. Such an approach is utilized in \cite{Ivanov:2000cm} in order to model the soft-hard interface of the gluon distribution. Here, we will assume that the role of this regulator is played by the saturation scale, $\mu=Q_s(x)$. Hence, it produces a cross section that behaves as $E\frac{d^3\sigma}{d^3p} \sim p_T^{-4}$ in the regime of high $p_T$, whereas corrections due to collinear radiation emission should conduct to $p_T^{-n}$ behavior, which will be embedded into the gluon distribution through the parameter $\delta n$. Such a quantity should grow in the hard region of the spectra. In the scaling region, one has $\phi(x,k_T^2)=\phi(\tau)$, where $\tau = k_T^2/Q_s^2(x)$. The unitarity of $S$ matrix, Eq. (\ref{eq:DISb}), will constrain the UGD normalization due to the fact that one should have $\int \frac{d\tau}{\tau} \phi(\tau)= \frac{3\sigma_0}{4 \pi^2 \alpha_s}$. Given these considerations, our ansatz for the gluon distribution is the following
\begin{eqnarray}
\label{eq:gluon}
\phi(x, k_T^2)=\frac{3\sigma_0}{4\pi^2\alpha_s}\frac{(1+\delta n)}{Q_s^2}\frac{k_T^2}{ \left(1+\frac{k_T^2}{Q_s^2} \right )^{(2+\delta n)}},
\end{eqnarray}
where  $Q_s$ and $\delta n$ dependencies on the energy have to be related to the growth of the total cross section as the collision energy increases. Using the paramatetrization above for $\phi(x,k_T)$ and considering central rapidity, the $p_T$ distribution of the produced gluons in Eq. (\ref{eq:fatkt}) is computed as
\begin{eqnarray}
\label{eq:sigmapt}
E\frac{d^3\sigma}{d^3p}&=& N_0\frac{\xi}{\xi -1}\left( 1-\frac{1+\xi\tau}{(1+\tau)^{\xi}}\right )\frac{1}{( 1+\tau)^{1+\xi}},\\
N_0 & = & \frac{9\sigma_0^2}{8C_F\pi^3\alpha_s},\,\,\xi=1+\delta n.
\end{eqnarray}

The rapidity distribution of the produced gluons and their mean momentum may be calculated by integrating Eq. (\ref{eq:fatkt}) over $p_T$ and $y$ without the necessity of a cutoff in the IR region,
\begin{eqnarray}
\label{eq:dsigmaY}
\frac{d\sigma}{dy}&=&\int d^2p_T \frac{d^3\sigma}{d^2p_Tdy}, \\
\left<\ p_T\right>&=&\frac{\int d^2p_T \frac{d^3\sigma}{d^2p_Tdy}p_T}{\int d^2p_T \frac{d^3\sigma}{d^2p_Tdy}}.
\end{eqnarray}

The total cross section, $\sigma_{tot}^{pp}(\sqrt{s})$,  calculation using the saturation formalism has been performed in \cite{Carvalho:2007cf}, where the authors split this quantity into two parts: $\sigma_T=\sigma_{sat}+\sigma_{pQCD}$, which corresponds to the contributions from the regions $\tau<1$ and $\tau>1$, respectively. The latter has been calculated using the QCD collinear factorization model. In this sense, the integral over $p_T^2$  in Eq. (\ref{eq:sigmapt}) produces the behaviors $Q_s^2(1+\delta n)^2$ for $\tau \ll 1$ and $Q_s^2/\delta n$ for $\tau \gg 1$, which indicates that for small values of $\delta n$ most part of the total cross section is due to the hard contribution towards the spectrum.

\begin{table*}[t] 
\centering
\caption{Parameters of the model for the QCD dipole cross section, determined from fits to data in the range $x \leq 0.08$ and $Q^2=[0.045, 10^4]$ GeV$^2$ \cite{Abt:2017nkc} (FIT A) and in the reduced range $Q^2=[0.01, 150]$ GeV$^2$ (FIT B) . Parameters for inclusive hadron production are also presented  (see the text for details).}
\label{tab:1}
\begin{tabular}{|c|c|c|c|c|c|c|c|}
\hline
                          & $\sigma_0$(mb) & $x_0$  $\times10^{-5}$    & $a$   & $b$ & $K$   & $\left< z \right>$ & $\frac{\chi^2}{\mathrm{dof}}$ \\ \hline
$\sigma^{\gamma *p} (FIT A)$      & 19.75 $\pm$ 0.09      & 5.05 $\pm$ 0.10 & 0.075 $\pm$ 0.002 & 0.188 $\pm$ 0.003    &       &                    & 2.48            \\ \hline
$\sigma^{\gamma *p} (FIT B)$      & 20.47 $\pm$ 0.61      & 3.52 $\pm$ 0.20 & 0.055 $\pm$ 0.039 & 0.204 $\pm$ 0.073    &       &                    & 1.74            \\ \hline
$pp \rightarrow \pi^0 +X$ &            &        &       &          & 1.361 $\pm$0.081 & 0.345 $\pm$0.006              & 1.50            \\ \hline
$pp \rightarrow h^\pm +X$ &            &        &       &          & 2.226 $\pm$ 0.065 & 0.418 $\pm$0.004              & 1.77            \\ \hline
\end{tabular}
\end{table*}

The inclusion of the hadronization process shall be performed analogously to the collinear factorization approach taking into account a hadron that carries a fraction $z$ of the gluon momentum,
\begin{equation}\label{eq:hadron}
\frac{d^3\sigma}{d^2p_{Th}dy}(pp\rightarrow h)=\int \frac{dz}{z^2}D_{g/h}(z,Q^2)\frac{d^3\sigma}{d^2p_Tdy}(pp\rightarrow g),
\end{equation}
where $z$ is the hadron momentum fraction, $p_{Th}=\frac{p_T}{z}$ is the gluon momentum, and $D_{g/h}(z,Q^2)$ is the gluon fragmentation function. The hadronization process might lead to the violation of the scaling once the fragmentation functions (FFs) depend on both $z$ and $Q^2$. In addition, the collinear FFs usually employed are valid from $Q^2>1$. As in Ref. \cite{Levin:Rezaeian}, we considered that the hadronization process can be approximated performing the substitution $p_T \rightarrow \frac{p_{Th}}{\left< z\right>}$. Also, it was supposed that $\left< z\right>$ does not vary within the scaling range. Thus, in this case one has
\begin{eqnarray}
\label{eq:hadron2}
\frac{d^3\sigma^{pp\rightarrow h}}{d^2p_{Th}dy}=\frac{K}{\left< z\right>^2}\frac{d^3\sigma^{pp\rightarrow g}}{d^2p_{Th}dy}\left(p_T=\frac{p_{Th}}{\left< z\right>}\right),
\end{eqnarray}

\begin{figure}[t]
\includegraphics[width=\linewidth]{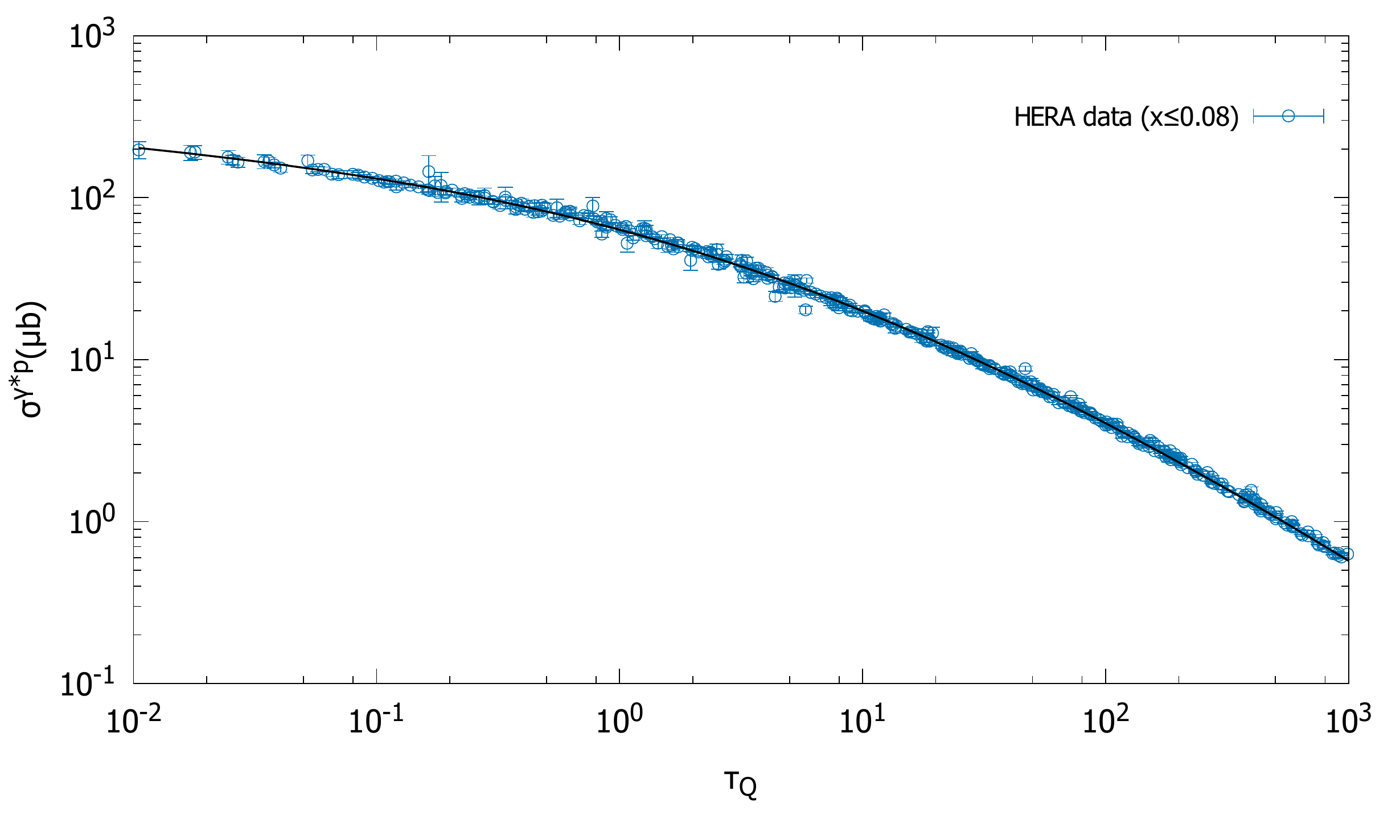}
\caption {Total cross section $\gamma^* p$ obtained from the dipole parametrization of Eq. (\ref{eq:DISc}) compared to data as a function of the scaling variable for the intervals $\tau_Q<10^3$ and $x\leq 0.08$.}
\label{fig:DIS} 
\end{figure}
where in Eq. (\ref{eq:hadron2}) we used the constants $K$ and $\left< z\right>$ to parametrize the hadronization process. It is important to mention that for $p_T<Q_s(x)$ the gluon spectrum approaches zero as $\tau^2$. However, the inclusion of fragmentation functions leads to an integration over $z$, Eq. (\ref{eq:hadron}), which continues to enhance towards the region of small $p_{Th}$. In Sec. \ref{sec:results}, both approaches are compared, and we also investigate the influence of FFs on hadron spectra. 

Having introduced the analytical expressions for the UGD and gluon/hadron invariant cross sections using Eqs. (\ref{eq:gluon}), (\ref{eq:sigmapt}), and  (\ref{eq:hadron2}), in the next section we determine the dependence on $\tau$ of the function $\delta n$  by an adjustment of the small-$x$ HERA data and then apply it to perform predictions for the charged and neutral hadron production cross sections.

\section{Results and discussions}
\label{sec:results}

Our procedure consists of fitting $\delta n $ from a total cross section of DIS within the scaling domain and then looking into how compatible it will be with the $p_T$ spectra of hadrons. The HERA data for $x<0.01$ supports geometric scaling whether they are plotted in terms of the ratio $Q^2/Q_s^2(x)$ with $Q_s(x)^2\sim x^{-\lambda}$, where $\lambda \sim 0.3$. We  assume that the same behavior is compatible with data of inclusive particle production in $pp(\bar p)$ collisions at high energies. This is clearly demonstrated in the studies of Refs. \cite{McLerran:2014apa,Praszalowicz:2011tc,Praszalowicz:2013fsa,Praszalowicz:2015dta}, where geometric scaling is shown to be present in $pp$, $pA$ and $AA$ collisions. In the present analysis, we have fixed $\lambda=0.33$ fitting $Q_s^2(x)$ and $\delta n$ from experimental data. Then, we have verified that data can be well described by supposing that $\delta n $ varies slowly within the whole scaling interval, being close to zero as $\tau \rightarrow 0 $ and $0.3$ for high $\tau$. Accordingly, this behavior has been modeled considering a powerlike form, which results in the following expressions
\begin{eqnarray}
\label{eq:pars}
 \delta n (\tau ) &=& a \tau ^b ,\\
 Q_s^2(x)&=& \left( \frac{x_0}{x}\right) ^{0.33}. 
\end{eqnarray}

Moreover, the QCD dipole cross section can  be  analytically computed  using the Fourier transform of the gluon distribution (\ref{eq:DISb}), which gives
\begin{eqnarray}
\label{eq:DISc}
\sigma_{q\bar q}(\tau_r)=\sigma_0\left (  1-\frac{2(\frac{\tau_r}{2})^{\xi}K_{\xi}(\tau_r)}{\Gamma(\xi)} \right ),
\end{eqnarray}
where $\tau_r = rQ_s(x)$ is the scaling variable in the position space, $\xi =1+\delta n$, and $\sigma_0$ is a free parameter related to the proton transverse area. We are left with 4 parameters  to be fitted, $\sigma_0$, $a$, $b$ and $x_0$. In what follows we present and discuss the results obtained by comparing the proposed parametrization with the total cross section data of DIS, Eq. (\ref{eq:DISa}). Afterwards, the  scaling property applied to the invariant  cross section for neutral pion and charged hadron production at different center-of-mass energies is looked into. In addition, there is also a discussion concerning the impact of saturation effects on inclusive gluon production and how it affects the hadron production at high energies. Moreover, the role played by the parameter $\delta n$ on observables regarding the IR domain is investigated. Specifically, we analyze the rapidity distribution of the produced gluons, which  is relevant for the inclusive total cross section determination within the saturation domain.

\begin{figure*}[t]
\includegraphics[width=0.9\textwidth]{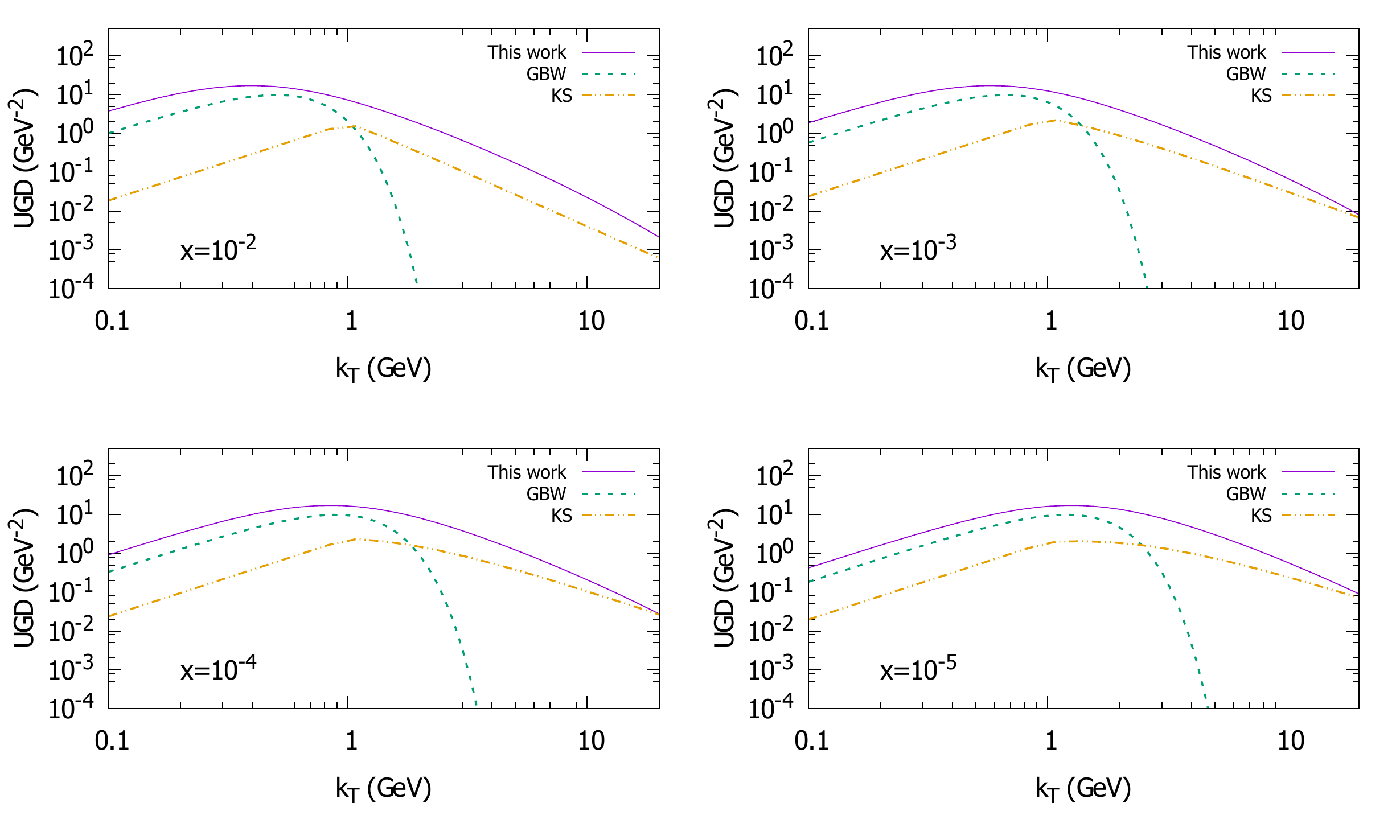}
\caption {Comparison of UGD obtained in this work with GBW \cite{GBWnovo} and KS \cite{Kutak:2012rf} at different values of $x$.}
\label{fig:UGD} 
\end{figure*}

In Table  \ref{tab:1} the fit results concerning HERA data \cite{Abt:2017nkc,Abramowicz:2015mha,Aaron:2009aa}  for $x \leq 0.045$ (FIT A) using the parametrization (\ref{eq:pars}) are presented. The Fig. \ref{fig:parametros} shows $\delta n(\tau)$  (left) in terms of the scaling variable $\tau_Q=Q^2/Q_s^2$ and $Q_s(x)$ (right) as a function of $x$ (considering $x$ and $\tau_Q$ ranges of experimental data). Regarding Fig. \ref{fig:DIS}, it displays the $\gamma^*p$ cross section as a function of the scaling variable $\tau_Q$. This quantity is determined using Eq. (\ref{eq:DISa}) along with the dipole cross section parametrization in Eq. (\ref{eq:DISc}). We can clearly see that $Q_s(x)=1$ GeV at $x_0=0.5 \times 10^{-4}$. These results are near to those encountered in analyses performed by \cite{GBWnovo}. The $\delta n$ parameter controls the cross section behavior for inclusive gluon production at $p_T>Q_s(x)$ and varies from $\delta n \sim 0.05$ at $\tau=0.01$ up to $\delta n=0.3$ at $\tau=10^3$. Such a fact implies that the cross section should depend on $p_T^{-4.6}$ at the edge of the region where scaling is broken, which is in agreement with the expoenent $n$ extracted from the cross section for jet production at high energies.
In Fig. \ref{fig:UGD} we compare the UGD obtained in this work with GBW parametrization \cite{GBWnovo} and KS \cite{Kutak:2012rf}, which reproduces DGLAP behavior at $k_T>Q_s(x)$. We can see that at low $k_T$  our parametrization behaves like GBW. However, at $k_T>Q_s$ the suppression presented by the Gaussian shape of the GBW distribution is too large to describe the hadronic spectra, where our parametrization gives close results  to the KS distribution that includes collinear resummation effects.

Here, some comments are in order. The main reason for including so high virtualities, $Q^2\sim 10^4$ GeV$^2$, is to cover the kinematic window $(\sqrt{s},\,p_T)$ of the measured hadron spectra. We are aware that this degradates the quality of fit. Therefore, a feasible reproduction of the measured data points at high $Q^2$ with a statistically acceptable confidence level was a
needed condition for interpreting the results. In addition, a fixed coupling constant has been considered, $\alpha_s=0.2$. This can be justified as most saturation approaches consider the average gluon (hadron) transverse momenta being of the order of the saturation scale in such way that $\alpha_s=\alpha_s(\langle p_T\rangle =Q_s^2)$, and eventually the logarithmic dependence on running coupling can be absorbed in the shape of the heuristic UGD. Indeed, this is the case by looking at Fig. \ref{fig:UGD} it becomes clear that the heuristic UGD reproduces the behavior of CCFM evolution equation (the linear contribution to the Kutak-Sapeta UGD) at large $k_T$.

From a theoretical point of view, we have no strong justification to extend the scaling fit to so high virtualities. In Ref. \cite{Iancu:2002tr}, it has been shown a long time ago that geometric scaling can be extended up to $Q^2\sim Q_s^4/\Lambda_{QCD}^2$. This gives $Q^2$ around 100 GeV$^2$ for $Q_s^2=2$ GeV$^2$. Moreover, in Ref. \cite{Kwiecinski:2002ep} it was demonstrated that geometric scaling is completely preserved by LO DGLAP evolution in the fixed coupling case (scaling violation by the contribution of the branch point singularity is a marginal effect at small-$x$). In the case of running coupling the scaling behavior gets violated, but it is possible to factor out the effect of such violation. Namely, the violation is proportional to the value of the $\alpha_s(Q^2=Q_s^2)$ evaluated at the saturation scale and the scaling is restored in the very same region proposed by \cite{Iancu:2002tr}. We reinforce that we are not pursuing an adjustment quality factor (QF). We have tested the fit in the range $x<0.01$ and $Q^2<150$ GeV$^2$, which produces a $\chi^2/dof \sim 1.7$ (FIT B, shown in Table  \ref{tab:1}). Despite having improved the QF, the corresponding range of validity in transverse momentum is substantially narrowed. Some concern about large-$x$ effects comes up and the role played by the quark-initiated processes should be addressed. Within the $k_T$-formalism, in Ref. \cite{Czech:2005vy} the leading-order diagrams involving quark degrees of freedom which are important in the fragmentation region were included. Indeed, the contribution at large $p_{Th}$ is sizable at low-energies (ISR) and at RHIC kinematic range (for $p_{Th} \gsim\, 3$ GeV). At the LHC those contributions are strongly suppressed and the dominant subprocess is $gg\rightarrow gg$ ($gg\rightarrow g$) in collinear factorization ($k_T$-factorization). Specifically, within the collinear factorization formalism at NLO accuracy it was demonstrated in Ref. \cite{Sassot:2010bh} (see Fig. 6 in that work) that gluon fusion subprocess dominates up to $p_{Th}\simeq 40$ GeV in $pp$ collisions at the LHC for $\sqrt{s}=7$ TeV.

\begin{figure*}[t]
\includegraphics[width=0.7\textwidth]{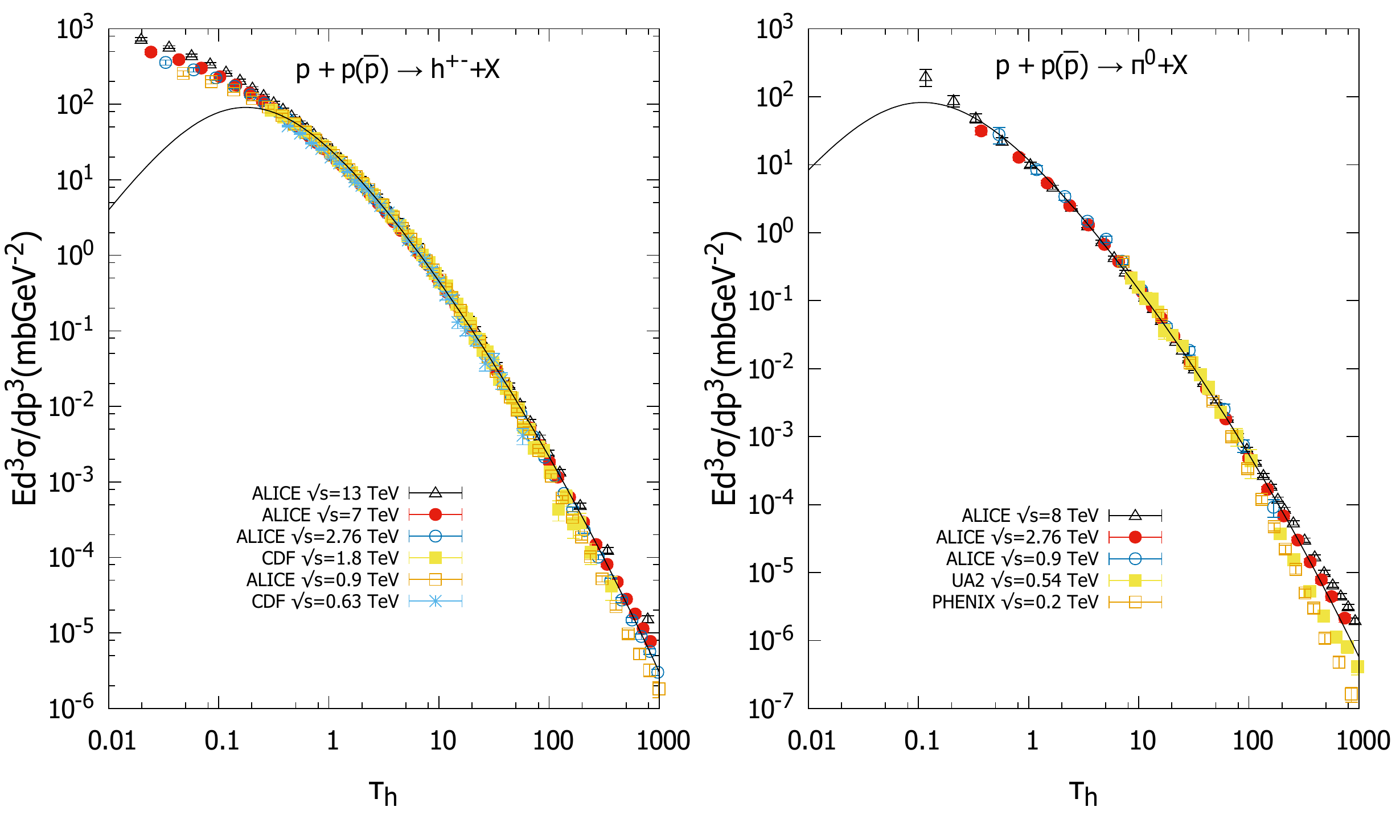}
\caption{Invariant  cross section in terms of the scaling variable, $\tau_h$, for charged hadron and neutral pion production at different values of centre-of-mass energies, $\sqrt{s}$. Extrapolation of fitting results (i.e., the range $1<\tau_h<100$ ) up to $\tau_h =10^3$ is  presented for sake of illustration.}\label{fig:escalonamento}
\end{figure*}

\begin{figure}[t]
\includegraphics[width=\linewidth]{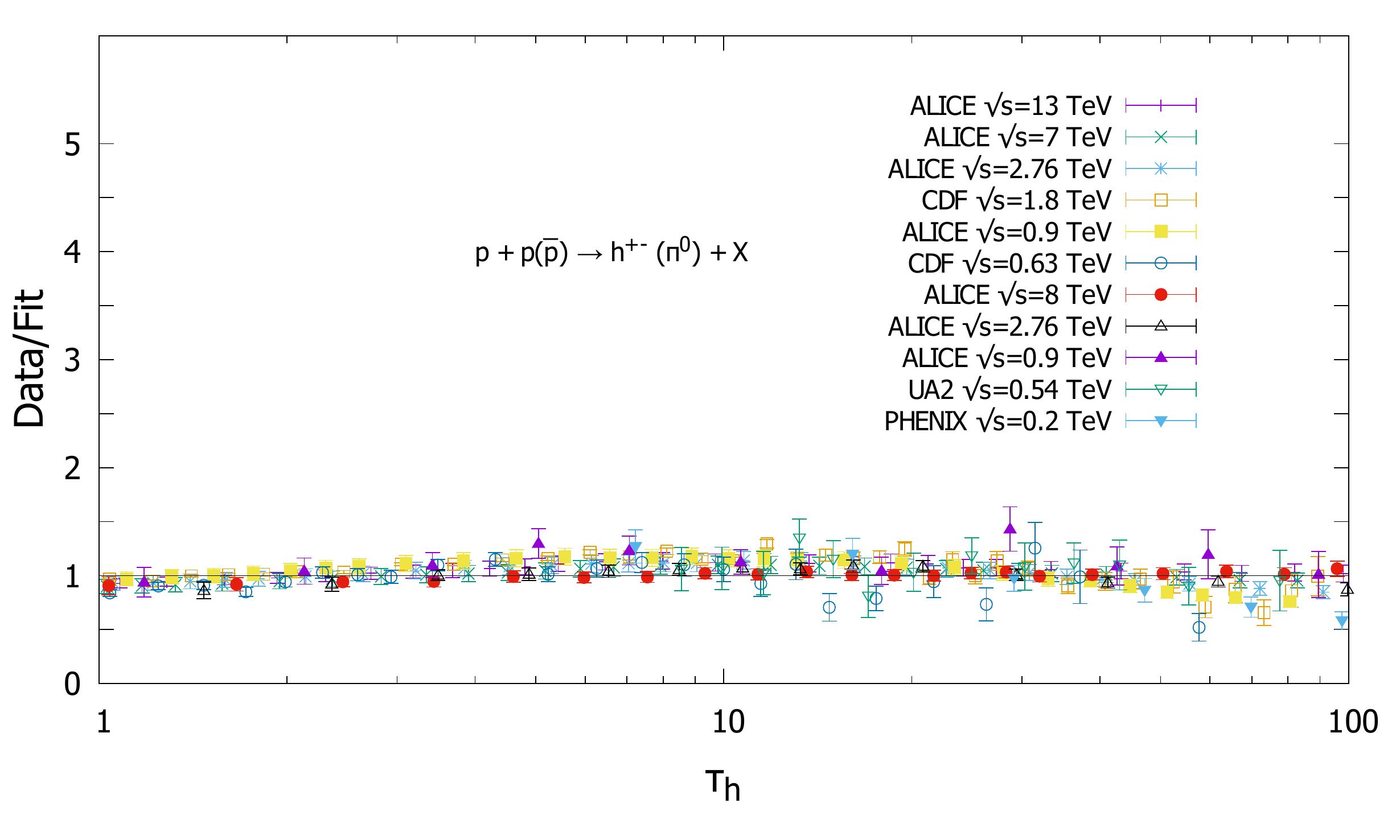}
\caption{The ratio data/theory as a function of the scaling variable, $\tau_h$, for charged hadron and neutral pion production in high energy colliders.}
\label{fig:ratiodatatheory}
\end{figure}

The scaling in inclusive hadron production in $pp$ and $p\bar{p}$ collisions becomes more evident whether we combine data from different colliders covering a large range of the scaling variable, $\tau_h$ and $x_h$. In such a case, we define $\tau_h=p_{Th}^2/Q_s^2(x_h)$ and $x_h=\frac{p_{Th}}{\sqrt s}$. Figure \ref{fig:escalonamento} presents data from ALICE and CDF collaborations \cite{ALICE13TEV, ALICE7TEV, Abe:1988yu} for charged hadron production (left) and data from ALICE, UA2 and PHENIX \cite{alicepi08, ALICEpi0276, ALICEpi0097, UA2l, PHENIXpi0200} for neutral pion production (right) compared to our prediction. The values of $\left<z\right>$ and $K$ were fitted within the range $1<\tau_h<100$ considering Eq. (\ref{eq:hadron2}), since for $\tau_h>100$ the scaling should be less accurate. It can be checked that the region $\tau_h<1$ of small $p_T$ is strongly sensitive to the hadronization process. This fact diminishes the accuracy of the gluon distribution scaling in this region of the hadronic spectra.  The fitted values are presented in Table \ref{tab:1} which considers data from $pp$ collisions, since we observed that in $p\bar p$ collision we have a significant amount of scaling violation. In this case the predictions to $p\bar p $ are obtained by extrapolation. The mean values of the momentum fraction $z$ carried by the hadron are fairly distinct in both cases, though they are close to those obtained from models that use collinear factorization \cite{Sassot:2010bh}. The data displayed in Fig. \ref{fig:escalonamento} covers the region $0.01<\tau_h<1000$ and may be related to $\tau$ whether it is taken into account that if $x=\frac{p_{T}}{\sqrt{s}}$ and $p_T=\frac{p_{Th}}{\left<z\right>}$, one gets $\tau_h=\tau  \left<z\right>^{2.33}$. The scaling domain extends up to $\tau=10^3$ for HERA data, which corresponds to $\tau_{\pi^0}=84$ and $\tau_{h^{\pm}}=131$. Besides, the scaling extension for higher values of $\tau_h$ observed in charged hadron spectra with relation to neutral pion can be understood by the difference in $\left< z\right>$. In Fig. \ref{fig:ratiodatatheory}, it shows the ratio theory/data as a function of the scaling variable $\tau_h$, for the sake of clarification about the effective quality of the fit.  Figure \ref{fig:sigmaPT} presents our results compared to data as a function of $p_{Th}$ within the scaling region for distinct values of $\sqrt{s}$. It can be noticed that the saturation effect leads to correct growth of  the spectra in terms of the enhancement of $\sqrt{s}$. It is important to realize that collinear factorization formalism requires some additional mechanism to reproduce the growth of cross section as a function of $\sqrt{s}$ through the factor $K(\sqrt{s})$ or by including the intrinsic transverse momentum.

The region of small $p_{Th}$ should be sensitive to the specific form of fragmentation functions and the hadron mass. Furthermore, there is a deviation of cross section scaling in this region. To measure the impact of FF on our results we calculated the cross section using KKP \cite{kkp} and HKNS \cite{HKNS} fragmentation functions in LO by integrating Eq. (\ref{eq:hadron}) and compared the result with the one where a constant value for $z$ is considered.  The momentum scale $Q^2$ utilized in the analysis is the hadron transverse momentum, $p_{Th}$. In the region where $Q^2<Q_0^2$ the scale is fixed at this value. The result is presented in Fig. \ref{fig:sigmaPTkkp} for $\pi^0$ production, and it can verified  that for $\tau_h<1$ there is a decrease in the growth of the spectra due to saturation of gluon distribution. On the other hand, the growth caused in this region is still higher than the one shown by data. The gluon FF is parametrized as $z^{\alpha}(1-z)^\beta$ and the enhancement related to the cross section within the region $\tau_h<1$ may be assigned to the parameter $\alpha$, which is considerably different in the case of KKP and HKNS fragmentation functions.

\begin{figure}[t]
\includegraphics[width=\linewidth]{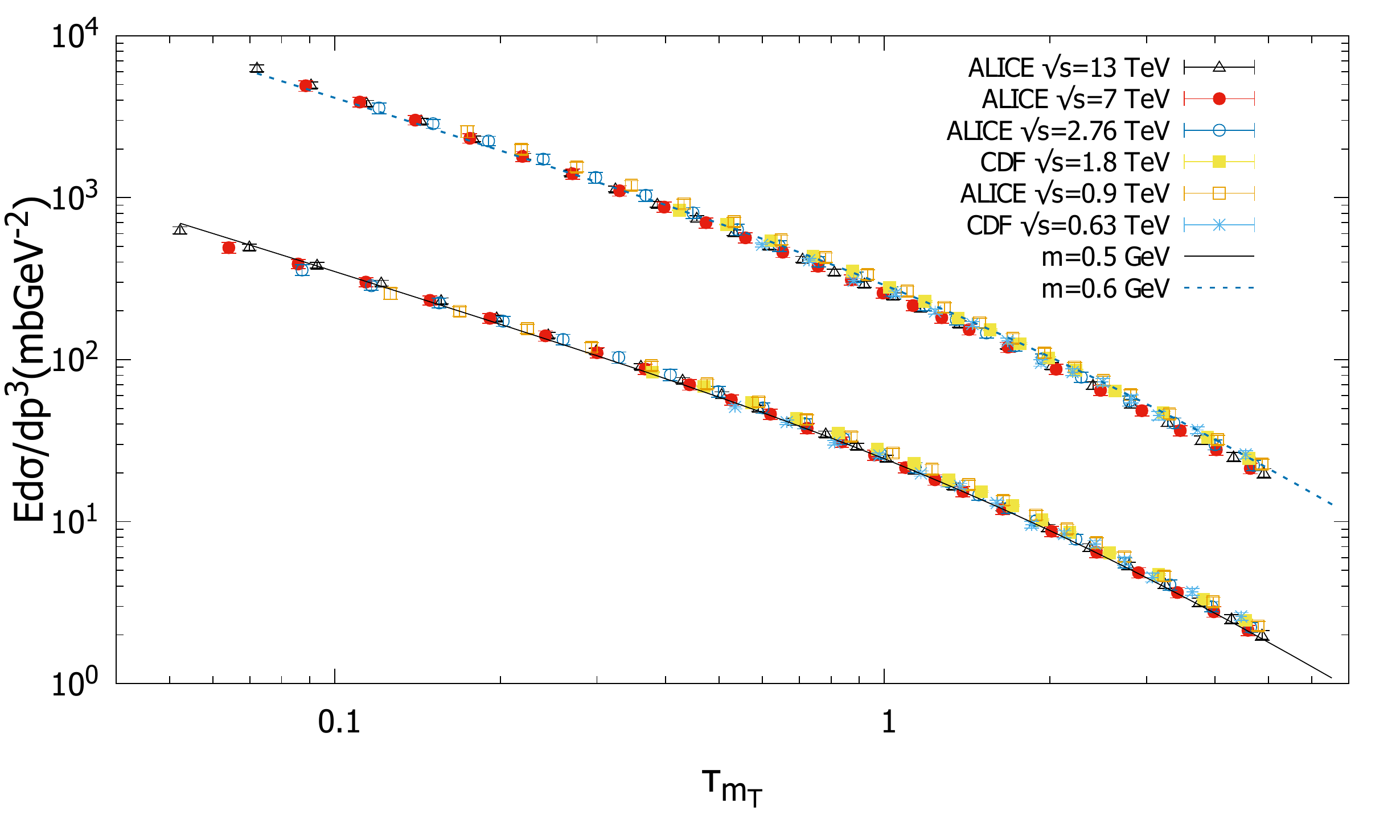}
\caption{The invariant cross section for charged hadrons as a function of the scaling variable, $\tau_{m_T}=m_T^2/Q_s^2$ in high energy colliders (see discussion in text).}
\label{fig:mtscaling}
\end{figure}

\begin{figure*}[t]
\includegraphics[width=0.7\linewidth]{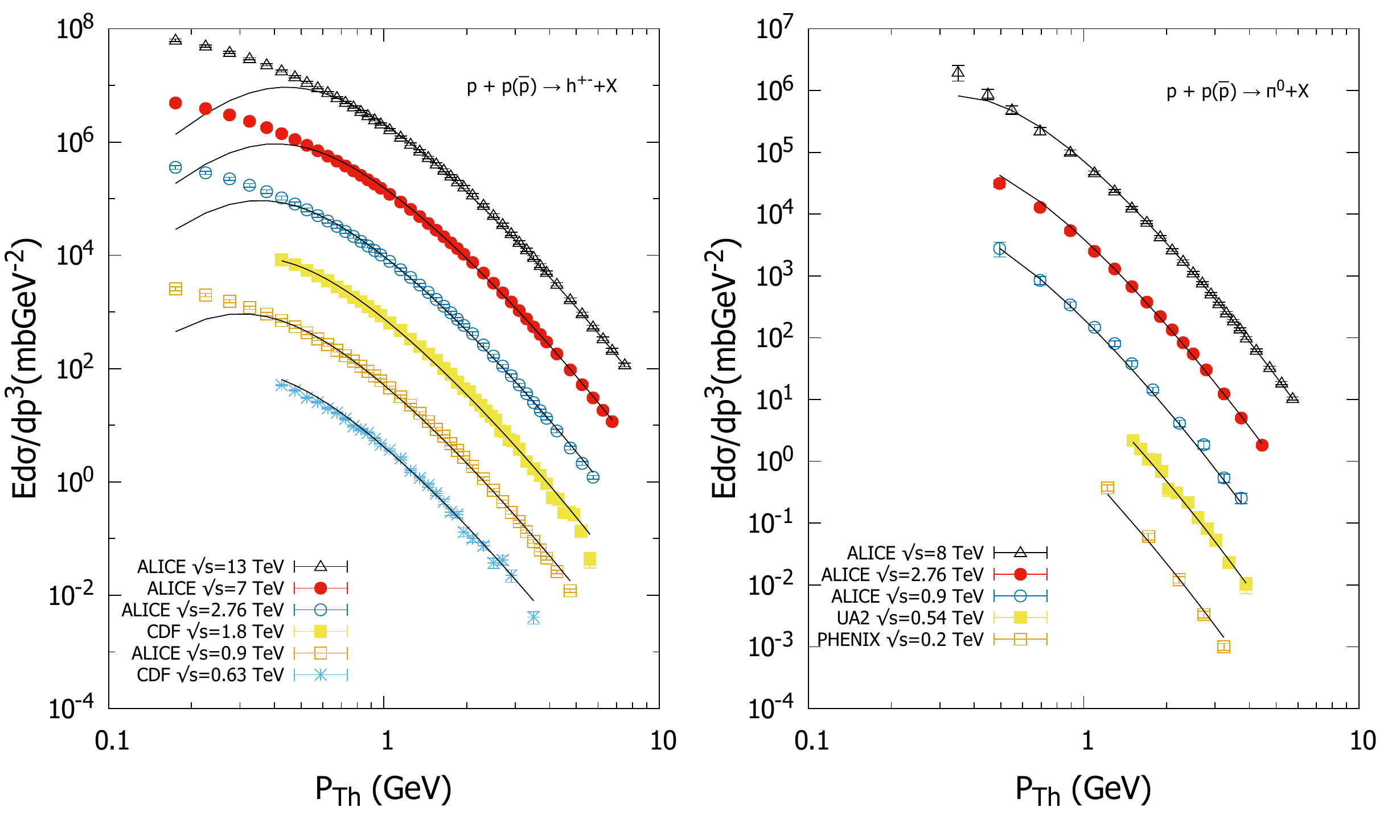}
\caption{Transverse momentum spectra within the scaling region $\tau<10^3$ as a function of $p_{Th}$. The data are multiplied by a factor $10^i$ at each energy for better visualization.}\label{fig:sigmaPT}
\end{figure*}

\begin{figure}[h]
\includegraphics[width=\linewidth]{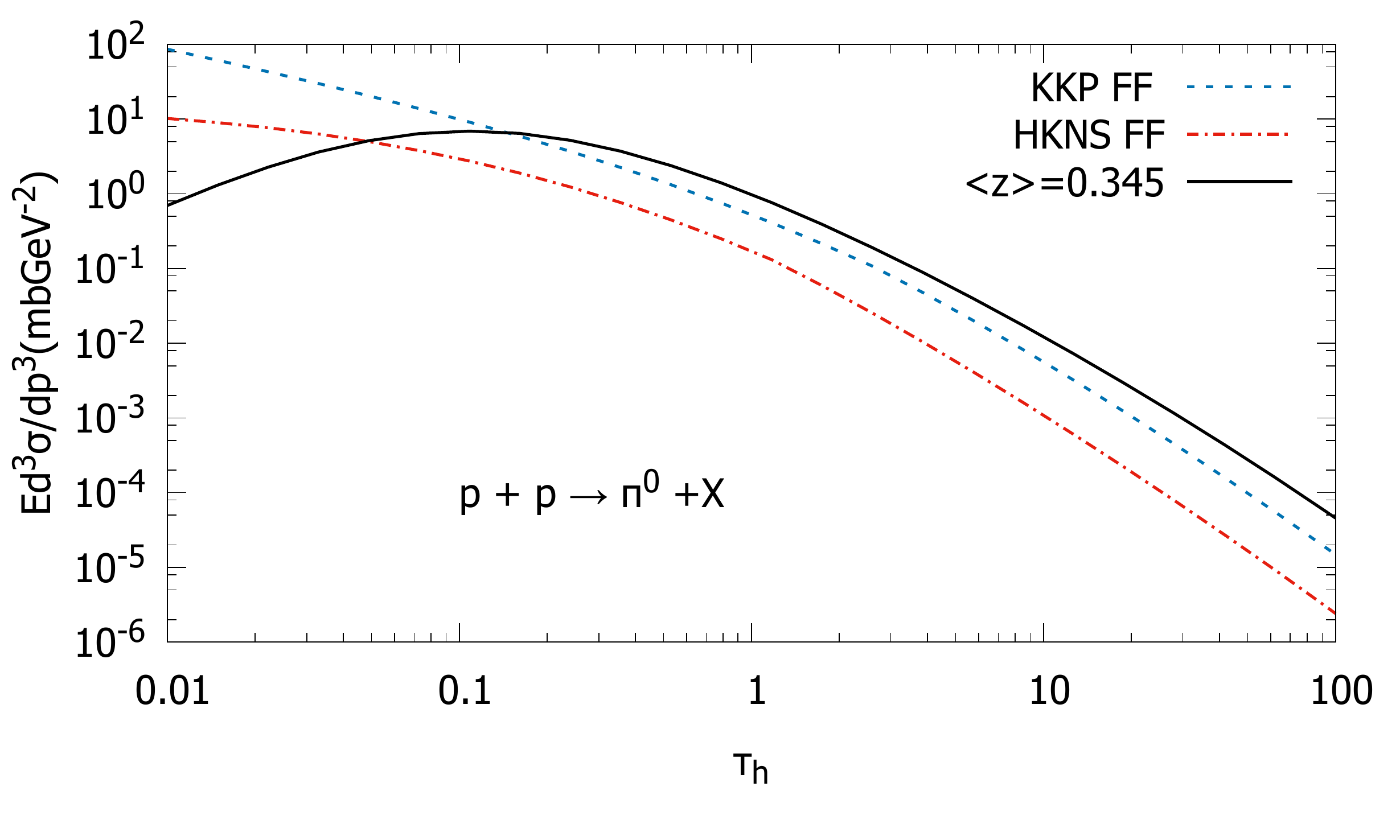}
\caption{Comparison between the results using $\left< z \right>$ and the fragmentation functions KKP and HKNS.} \label{fig:sigmaPTkkp} 
\end{figure}

We address the low-$p_T$ behavior in an exploratory study. We follow Ref. \cite{Levin:Rezaeian} closely where the gluon transverse momentum, $p_T$ is replaced by $p_T^2\rightarrow p_T^2 +m_{\mathrm{jet}}^2$ with $m_{\mathrm{jet}}$ being an effective minijet mass. This procedure naturally regulates the denominator in Eq. (\ref{eq:fatkt}) due to the presence of a nonzero jet mass. The value of minijet mass is considered to be proportional to the saturation scale, $m_{\mathrm{jet}}^2\sim 2\mu_{np} Q_s$, where $\mu_{np}$ is the scale of the soft interactions. For instance, the typical value of the  saturation scale at central rapidities for 13 TeV and $p_T\approx 1$ GeV is $Q_s(\sqrt{s},p_T)\simeq 0.93$ GeV. Moreover, we can consider the soft scale being of order $\mu_{np}\sim \Lambda_{QCD}\simeq 0.3$ GeV. This will give for low $p_{Th}$ charged hadrons at the LHC, $m_{\mathrm{jet}}\approx 2\Lambda_{QCD} Q_s(\eta=0)\simeq 0.56$ GeV. We now make the hypothesis that the new scaling variable is $\tau_{m_T}=m_T^2/Q_s^2$ with $m_T = \sqrt{p_{Th}^2+m_{\mathrm{jet}}^2\langle z\rangle^2 }$. In Fig. \ref{fig:mtscaling} we compare the invariant cross section for charged hadrons as a function of $\tau_{m_T}$ for two fixed values of minijet mass, $m_{\mathrm{jet}}=0.5$ (solid line) and 0.6 GeV (dashed line) in order to estimate the effect of the effective jet mass. We used the same parameters for the charged hadrons fit, with $\langle z \rangle = 0.418$. For better visualization, the results for $m_{\mathrm{jet}}=0.6$ GeV were multiplied by a factor of 10. The quality of data description at low $p_T$ is somewhat reasonable. Interestingly enough, recently the low-$p_T$ region is assumed to be dominated by a thermal contribution and the hard scattering contribution does not play a significant role there. This has been investigated, for instance in Refs. \cite{Baker:2017wtt,PhysRevD.100.034013,Gotsman:2019vrv}. 

\begin{figure*}[t]
\includegraphics[width=0.7\linewidth]{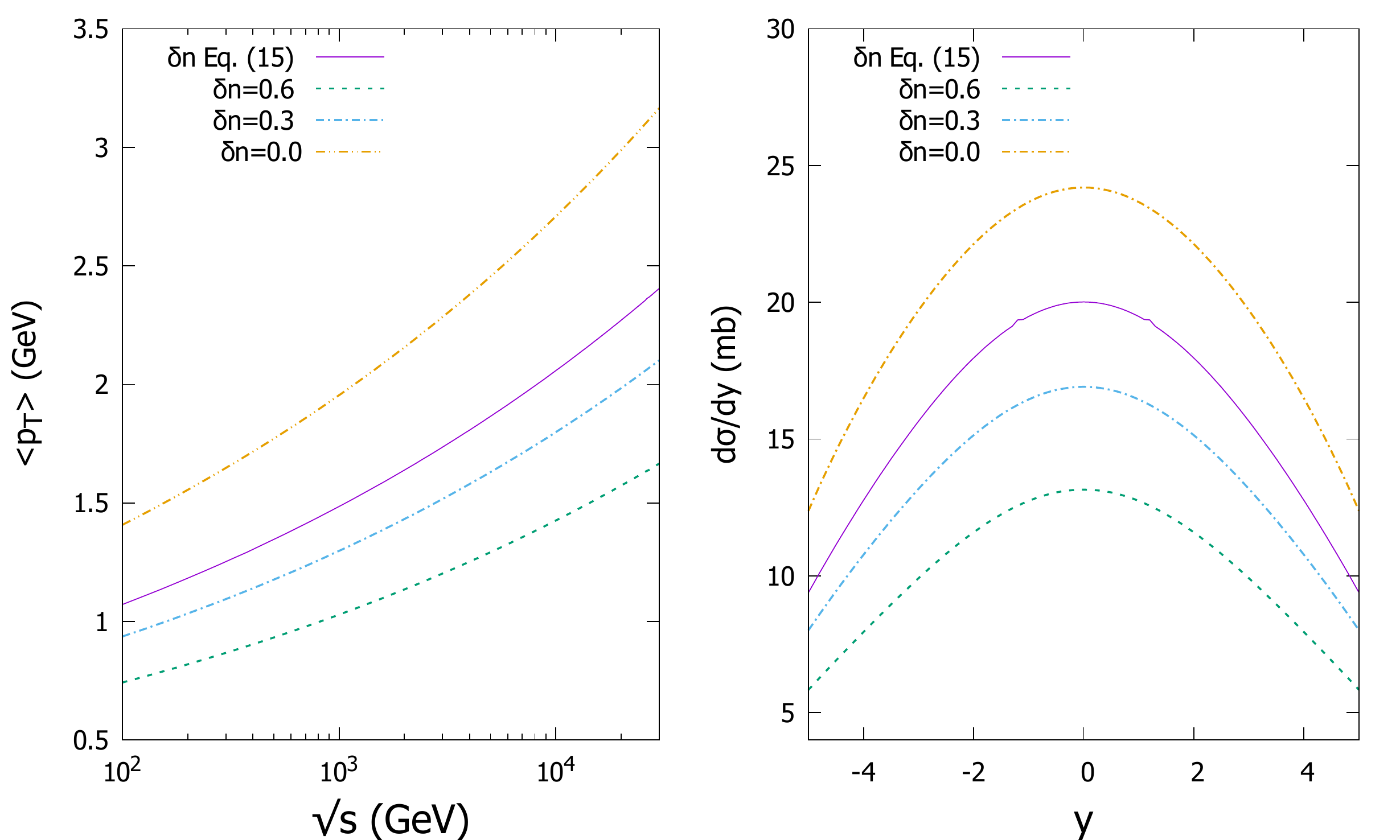}
\caption{Rapidity distribution at $\sqrt s=13$ TeV and mean transverse momentum of the produced gluons for distinct values of $\delta n$.}\label{fig:ptmedio}
\end{figure*}

Finally, we have estimated the rapidity distribution of the produced gluons by integrating Eq. (\ref{eq:dsigmaY}) over $p_T$ for $\sqrt{s}=13 \ {TeV}$ at the LHC considering different values of $\delta n$, which are shown in Fig. \ref{fig:ptmedio}. We observe that even a small increase of $\delta n$ implies on a significant reduction of the cross section. In particular, the variation of $\delta n $ from $0$ up to $0.3$ leads to a decreasing of the distribution in the central region by a factor $2/3$. Figure \ref{fig:ptmedio} shows the mean values of gluon $p_T$ as a function of $\sqrt{s}$ for the same values of $\delta n$. Clearly, the highest values of $\delta n$ leads  to higher mean momentum of the produced gluon. In the case of the hadronic spectra, the calculation of $\left<p_{Th}\right>$ and $d\sigma/dy$ depends basically on the behavior of the fragmentation funtions within the region $p_{Th} < 1$. The data from the CMS collaboration \cite{CMSPTM} shows that $\left< p_{Th^{\pm}}\right>=0.5$ at $\sqrt{s}=2.76$ TeV. This is compatible with our results using $z\simeq 0.3$. There are different approaches for the calculation of rapidity distribution and mean transverse momentum of the produced hadrons that are strongly dependent on the region $p_T<1$, such as the inclusion of intrinsic momentum in Ref. \cite{Levin:Rezaeian} or the extension of FFs for this region \cite{Tribedy}.

The analysis presented here is very close to the ones discussed  in Refs. \cite{McLerran:2014apa,Praszalowicz:2011tc,Praszalowicz:2013fsa,Praszalowicz:2015dta}.  There, the invariant cross section is written in terms of a universal function, $F(\tau_h)$, which is modeled phenomenologically making use of the Tsallis parametrization \cite{Tsallis:1987eu},
\begin{eqnarray}
\label{tsallis}
\frac{d\sigma (pp\rightarrow h)}{dyd^2p_T}&=& \frac{S_{\perp}}{2\pi}F_h(\tau_h),\\
\label{ts2}
F_h(\tau_h) & \approx & N_h\left[1+\frac{\tau_h^{1/(2+\lambda)}}{n_h\kappa_h}  \right]^{-n_h},
\end{eqnarray}
where the Tsallis temperature of the hadron of specie $h$ is given by $T_h\approx \kappa_h\langle Q_s(x)\rangle $ \cite{Praszalowicz:2013fsa}. As the temperature is driven by the average saturation scale, $\langle Q_s\rangle$, it is energy dependent. The constant $\kappa_h\sim 0.1$ is fitted from thermal distributions of hadrons. The overall normalization, $N_h=\gamma_hb_h/ (2\kappa_h^2)$, depends on the constants $\gamma_h$ and $b_h$ which can be calculated analytically in terms of $\kappa_h$ and the energy independent constant, $n_h$ \cite{Praszalowicz:2013fsa}. Comparing our expression in Eq. (\ref{eq:sigmapt}), we identify $n_h\sim (3+2\delta n)$. However, the functional form of $F(\tau_h)$ in the present work is quite distinct of that in \cite{Praszalowicz:2013fsa}.

The very same Tsallis-like parametrization described above is investigated in Ref. \cite{Rybczynski:2012vj}, where the scaling function is given by
\begin{eqnarray}
\label{Osada}
F_h(\tau_h) = \left[1+(q-1)\frac{\tau_h^{1/(2+\lambda)}}{\kappa}  \right]^{-1/(q-1)},
\end{eqnarray}
where the nonextensive parameter $q=1.134$ and $\kappa=0.1293$ have been determined recently \cite{Osada:2017oxe} using the available data on inclusive hadron production. The description of geometric scaling in the semi-inclusive transverse momentum spectra in $pp$ collisions taking into account the same formalism has been done in Ref. \cite{Osada:2019oor}. In  \cite{Osada:2019oor} the  inclusive distribution with fixed multiplicity or limited multiplicity class is considered and it is assumed the same relations (\ref{tsallis}) and (\ref{Osada}) as for inclusive case. In addition, the replacements $S_{\perp}\rightarrow S_{\perp}^*$ and $\sqrt{s}\rightarrow \sqrt{s}^*$  are performed, where the latter is the effective energy replacing the actual colliding energy.

Still along the Tsallis-like distribution, in Refs. \cite{Zhang:2014dna,Yang:2017cup} scaling is also observed for the variable $z=p_T/K$ with  $K$ being a scaling parameter energy dependent. The scaling function is related to the $p_T$-spectra in the form $\Phi_h(z) =AE\frac{d^3\sigma}{d^3\vec{p}}(p_T=Kz)$, where  the parameters $K$ and $A$ depend on the collision energy. The scaling for identified hadrons, $\Phi_h$ (with $h=\pi,\,K,\,p$), has the following form
\begin{eqnarray}
\label{Zhang}
\Phi_h(z) = \left[1-(1-q)\frac{\sqrt{m_h^2+z^2}-m_h}{z_0}  \right]^{1/(1-q)},
\end{eqnarray}
where $C_q$, $q$ and $z_0$ are free parameters, $m_h$ is the mass of the particle species, and ($1-q$) is a measure of the nonextensivity  \cite{Zhang:2014dna}. The formalism has been extended in \cite{Yang:2017cup} in order to include the scaling behavior in the $p_T$ spectra of strange particles ($K^0_S,\,\Lambda,\,\Xi,\,\phi$) at $pp$ high energy collisions.

In a related study, in Ref. \cite{Petrovici:2018mpq} the average transverse momentum, $\langle p_T \rangle$, dependence of identified light flavor charged hadrons on the quantity $\tau_n=\sqrt{\left(\frac{dN}{dy}\right)/S_{\perp}}$ has been investigated.  Local parton-hadron duality and dimensionality arguments foresee the depletion of the ratio between the mean transverse momentum
and the square root of the hadron multiplicity per unit of rapidity and unit of the colliding hadrons transverse overlapping area towards central collisions. Namely, $R_h= \langle p_T\rangle/  \sqrt{\left(\frac{dN}{dy}\right)/S_{\perp}}$ is proportional to $1/(n\sqrt{n})$ where $n$ is the number of charged hadrons produced via gluon fragmentation. In that work, the scaling variable is the  quantity $\tau_n$ and it is expected that the global properties of the hadron production are determined by the properties
of flux tubes of size $\sim 1/\tau_n$ and are weakly influenced by the size of the colliding system.  The slopes of the average $p_T$, $\langle p_T\rangle$,  particle mass dependence and the $\langle \beta_T \rangle$ parameter from Boltzmann-Gibbs Blast Wave (BGBW) fits scale nicely with $\tau_n$. The successfulness of the scaling parametrizations above for the  single-particle distribution from the statistical mechanics point of view is based on a data description using only 3 degrees of freedom. Namely,  in the lowest-order approximation the production process is characterized by a power index $n$ which can be represented by a nonextensivity parameter $q=(n + 1)/n$, the average transverse mass, $\langle m_T \rangle$, and the overall normalization $A$ that is related to the
multiplicity per unit rapidity when integrated over transverse momentum. In particular, the average transverse mass can be represented by an effective temperature $T=\langle m_T \rangle/n$. For instance, in Fig. 8 of Ref. \cite{Wong:2015mba} both the low and large $p_T$ single hadron spectra is adequately described (compared to data from UA1, ALICE, ATLAS and CMS) by using the simple nonextensive statistics parametrization,
\begin{eqnarray}
\left. E\frac{d^3\sigma (pp\rightarrow h)}{d^3\vec{p}}\right|_{y=0} & = & Ae_q^{-m_T/T},\\
e_q^{-m_T/T} & = & \left[1+\frac{m_T}{nT}  \right]^{-n},\,n=\frac{1}{(q-1)}, \nonumber
\end{eqnarray}
where $m_T=\sqrt{m_{\pi}^2+p_T^2}$. The quality of the data description was subsequently corroborated by a series of similar works \cite{Parvan:2016rln,Grigoryan:2017gcg,Bhattacharyya:2017cdk,Shen:2019zgi}.

It was argued in \cite{Wong:2015mba} that the simplification of all complicated stochastic dynamics in hard scattering can be considered as a {\it no hair} reduction from the microscopic description to nonextensive statistical mechanics \cite{Tsallis:2017fhh}. Therefore, the inherent complexities at microscopic level disappear and are subsumed behind the stochastic processes and integrations. Interestingly, it has been recently proposed \cite{Deppman:2016fxs,Deppman:2019klo} that  fractal structures  cause the emergence of non extensivity in the system described by Tsallis statistics. The thermodynamical aspects of such a system are connected to the microscopic interaction of its pieces through the S-matrix.

\section{Summary and conclusions}
\label{sec:conc}

In this work we have investigated the role played by the geometric scaling for inclusive hadron production at high energies taking into account a phenomenological parametrization for unintegrated gluon distribution function. Also, we have proven that scaling is a good approximation within a large interval of $p_{Th}$ and $\sqrt{s}$. The decreasing related to the growth of total cross section at small $p_T$ may be viewed as an effect due to the saturation of gluon production in that region. Moreover, we have showed that the saturation formalism applied to a dipole cross section produces the correct growth of the spectra concerning the produced hadrons in $pp$ collisions as the energy increases. In this case, we have evidenced that the enhancement of the spectra in terms of $\sqrt{s}$ and the power index $p_T^n$ that describes the behavior of this observable in the region of high $p_T$ are related through the saturation scale. In the region where $\tau_h<1$, it can be seen that there is a huge influence of the hadronization process from the produced gluons. In this context, some mechanism for the soft hadronization is somewhat necessary. Furthermore, we have verified that the behavior of gluon distribution in the region of high $p_T$ has a strong impact on the determination of the rapidity distribution and inelastic total cross section, once within the saturation formalism the soft region is regulated by the scale $Q_s(x)$. Such a fact implies that these quantities receive an important contribution from this region.

\section*{Acknowledgments}

This work was financed by the Brazilian funding
agencies CNPq and  CAPES.

\bibliographystyle{h-physrev}
\bibliography{referencias_pp}

\end{document}